# Overcoming Intrinsic Dispersion Locking for Achieving Spatio-Spectral Selectivity with Misaligned Bi-metagratings

Ze-Peng Zhuang[1,2]†, Xin Zhou[1]†, Hao-Long Zeng[1]†, Meng-Yu Li[1], Ze-Ming Chen[1], Xin-Tao He[1], Xiao-Dong Chen[1], Lei Zhou[2]*, Jian-Wen Dong[1]*

**Affiliations:**

[1]State Key Laboratory of Optoelectronic Materials and Technologies and School of Physics, Sun Yat-sen University; Guangzhou, 510275, China.

[2]State Key Laboratory of Surface Physics, Key Laboratory of Micro and Nano Photonic Structures (Ministry of Education), Shanghai Key Laboratory of Metasurfaces for Light Manipulation and Department of Physics, Fudan University; Shanghai, 200438, China.

†These authors contributed equally to this work

*Corresponding author. Email: phzhou@fudan.edu.cn, dongjwen@mail.sysu.edu.cn

**Abstract:** Spatio-spectral selectivity, the capability to select a single mode with a specific wavevector (angle) and wavelength, is imperative for light emission and imaging. *Continuous* band dispersion of a conventional periodic structure, however, sets up an intrinsic locking between wavevectors and wavelengths of photonic modes, making it difficult to *single* out just one mode. Here, we show that the radiation asymmetry of a photonic mode can be explored to tailor the transmission/reflection properties of a photonic structure, based on Fano interferences between the mode and the background. In particular, we find that a photonic system supporting a band dispersion with certain angle-dependent radiation-directionality can exhibit Fano-like perfect reflection at a single frequency and a single incident angle, thus overcoming the dispersion locking and enabling the desired spatio-spectral selectivity. We present a phase diagram to guide designing angle-controlled radiation-directionality and experimentally demonstrate double narrow Fano-like reflection in angular (±5°) and wavelength (14 nm) bandwidths, along with high-contrast spatio-spectral selective imaging, using a misaligned bilayer metagrating with tens-of-nanometer-scale thin spacer. Our scheme promises new opportunities in applications in directional thermal emission, nonlocal beam shaping, augmented reality, precision bilayer nanofabrication, and biological spectroscopy.

# 1 Introduction

Wavelength and wavevector (propagation angle) are two fundamental properties of light. The ability to single out a photonic mode, exhibiting a specific wavelength and propagation angle, is crucial for many applications ranging from thermal emission [1,2] to high-contrast imaging [3,4] and high-beam-quality lasing [5,6]. In the past decades, wavelength-selectivity of light has been enabled by photonic crystals[7-11], whispering-gallery microcavities[12,13] and plasmonic lattices[14,15], while angle-selectivity of light has been realized by another group of structures/effects such as zero-index structured materials [1,16,17] and Brewster angle-enabled effects [18-20]. However, filtering out just one single mode with both desired wavelength and travelling angle, usually called as the "spatio-spectral selectivity", is hard to realize with a simple device. Although a one-dimensional photonic crystal was proposed to achieve such effect [21], experimental realization of the device was impeded by complex fabrication processes involving numerous multilayer films (~ 5λ), not mentioning its bulky size being highly unfavorable for optical integration. Meanwhile, designing optical gratings with tailored high-order diffractions present an alternative way, but the underlying physics is buried inside sophisticated optimization processes, which also inevitably introduce crosstalk between diffractions of different orders [22,23]. Without an intuitive model as a guidance, it remains a great challenge to achieve spatio-spectral selectivity with a thin photonic structure. This difficulty roots in the inherent dispersion that links wavelength and wavevector in periodic photonic structures, which results in *continuous* angular resonances.

Fano resonance can generate unique frequency-selective line-shapes for photonic structures, serving as an ideal mechanism to achieve the spectra-selectivity. Such an intriguing effect arises from the interference between a single mode with a high quality (Q) factor and a continuum of modes[24,25]. By tailoring the couplings between adjacent nanostructures [26,27] and the multiple-scattering effects in multilayer films [28-31] and subwavelength gratings [32,33], one can control the central wavelength and linewidth of the Fano resonance, thereby enabling customized spectral selectivity. In particular, a bound state in the continuum (BIC) can appear as the Fano resonance in a periodic photonic structure is "designed" to exhibit a zero bandwidth, which exhibits topological robustness characterized by an integer topological charge [34-37]. Perturbation around the BIC, however, can induce spectral selectivity with central-wavelength and linewidth dictated by the nature of perturbation[38-42]. On the other hand, varying the angle of incident light can excite high-Q modes with *different* (tangential) wavevectors, thereby shifting the central wavelength of Fano resonance and allowing for angle-sensitive multi-function [43,44] and sensing [45,46]. However, since wavelengths and wavevectors of photonic modes are strictly locked in a continuous band of a conventional periodic structure, such a mechanism fails to realize the desired spatio-spectral selectivity, as a mode can always be picked up no matter how incident angle varies.

Here, we reveal that, the radiation asymmetry of the high-Q mode in a photonic structure, often overlooked in previous studies, can break the wavelength-angle locking of Fano line-shapes leading to the desired spatio-spectral selectivity. Since the Fano line-shape is generated by the interference between scatterings from the high-Q mode and the background, modulating the radiation-directionality of the mode with varying incident angles can naturally change the Fano-resonance intensity, which helps bypassing the dispersion-induced limitation. We present a complete phase diagram for reflection dip ($R_{dip}$) and peak ($R_{peak}$) of Fano resonance versus the radiation-directionality of the photonic mode, which reveals that a mode with P-symmetric radiation can ensure zero $R_{dip}$ while the radiation-directionality of the mode dictates $R_{peak}$. We design a set of misaligned bilayer metagratings, and numerically demonstrate that the inter-layer offset is a crucial parameter to control the radiation-directionality of the guided-resonance mode, assisted with the near-field coupling between two layers with fixed inter-layer spacer. Guided by these phase diagrams, we finally design and fabricate a bilayer metagrating with inter-layer offset 37 nm and inter-layer distance 35 nm, and employ angle-resolved cross-polarization measurements to demonstrate that its allows for spatio-spectral-selective reflection at the 1349 nm wavelength and 0° incident angle, exhibiting a bandwidth of 14 nm for wavelength and ±5° for incident angle. Such spatio-spectral selectivity can yield many applications in practice, and we experimentally demonstrate a high-(wavelength-angle)-contrast imaging using our device as an explicit example.

## 2 Results

### 2.1 Phase diagram for spatio-spectral selectivity

We start from discussing how to realize spatio-spectral selectivity with a thin device based on a conceptual model. Obviously, a homogeneous slab does not exhibit spatio-spectral selectivity, as it allows light transmission in a wide range of incident angle and wavelength, as shown in Fig. 1a. As certain resonances (e.g., guided mode resonances) are induced in such system with periodic structuring, the resulting optical grating can exhibit a Fano-like reflection spectrum with high reflection occurring at a specific wavelength, thereby enabling the desired spectral selectivity. However, the band dispersion in a periodic system locks wavelength and (tangential) wavevector of the resonance mode under consideration, indicating that the wavelength selected by the Fano effect must shift with the incident angle, as depicted in Fig. 1b.

How to break such wavelength-angle locking to pick up just a *single* mode from the *continuous* dispersion? Noting that it is $R_{\text{peak}}$ (see Fig. 2a) that ultimately dictates the mode-filtering performance, we find that a specific mode can be effectively singled out if we can maintain a high $R_{\text{peak}}$ for that particular mode but suppress $R_{\text{peak}}$ for all other modes, without altering the inherent dispersion relation. Obviously, a conventional grating exhibiting mirror symmetry does not fulfill the above requirement. According to the coupled-mode-theory (CMT), reflection ($r$) and transmission ($t$) coefficients of such a system are generally given by

$$r = \tilde{r} + \frac{e^{j2\Theta}}{j(\omega\tau_0 - \omega_0\tau_0)+1}, \quad t = \tilde{t} + \frac{e^{j2\Theta}}{j(\omega\tau_0 - \omega_0\tau_0)+1} \tag{1}$$

where $\{\tilde{r},\tilde{t}\}$ represent the reflection and transmission coefficients of the background medium, $\omega_0$ and $\tau_0$ denote, respectively, frequency and radiation damping of the guided-resonance mode supported by the metagrating, and $\Theta = -\arg[\tilde{r}+\tilde{t}]/2$ defines the phase of the excited mode dictated by the local field (Supplementary Section 1). Equation (1) guarantees that $r = 0$ and $t = 0$ (and thus $R_{\text{dip}} = 0$ and $R_{\text{peak}} = 1$) can *always* be satisfied at two separate frequencies. Such a conclusion is robust against the variation of incident angle, since the latter can only change the property (i.e., $\omega_0$ and $\tau_0$) of the mode but does not violate the mirror symmetry of the system (see Fig. S3a,b).

The above argument motives us to break the mirror symmetry of the system, so as to realize angle-dependent $R_{\text{peak}}$. According to the CMT, we find that a metagrating *without* mirror-symmetry but exhibiting the P-symmetry exhibits the following expressions for reflection and transmission coefficients

$$r = \tilde{r} + \sqrt{1-\eta^2}\frac{e^{j(\Theta_1+\Theta_2)}}{j(\omega\tau_0 - \omega_0\tau_0)+1}, \quad t = \tilde{t} + (1-\eta)\frac{e^{j2\Theta_1}}{j(\omega\tau_0 - \omega_0\tau_0)+1} \tag{2}$$

where $\{\Theta_1,\Theta_2\}$ and $\eta$ denote the radiation phases and directionality of guided-resonance modes (Supplementary Section 1). Compared to Eq. (1), here the asymmetry of metagrating induces guided-resonance modes with unequal radiation phases $\{\Theta_1,\Theta_2\}$ and amplitudes $\{\sqrt{1-\eta^2},(1-\eta)\}$ to reflection and transmission sides, which provides an additional degree of freedom to control $R_{dip}$ and $R_{peak}$ via interferences with the background signals. Specifically, we find that while $R_{dip} = 0$ can still be satisfied at certain frequency, $R_{peak} = 1$ can only be met as $\eta = 0$ (Supplementary Section 1 and Fig. S3c,d). Therefore, if we design a P-symmetric metagrating to exhibit guided-resonance modes with angle-dependent radiation-directionality $\eta(\theta)$, we can efficiently modulate $R_{peak}$ as a function of $\theta$, thus picking up a single mode at the incident angle with $\eta = 0$. In this way, we can achieve the desired spatio-spectral selectivity at a pre-designed incident angle (Fig. 1c), through "designing" the function of $\eta(\theta)$.

More generally, we can derive analytical formulas for $R_{peak}$ and $R_{dip}$ in terms of radiation directionality of the guided-resonance mode. Considering the time-reversal symmetry of the wave scattering of our system under an oblique angle incidence [47,48] (Supplementary Section 2 and Fig. S1), we find from the CMT that

$$R_{peak} = \frac{\left[\sqrt{(1+\eta_{-k_\parallel})(1+\eta_{k_\parallel})} + \sqrt{(1-\eta_{-k_\parallel})(1-\eta_{k_\parallel})}\right]^2}{4}$$

$$R_{dip} = \frac{\left[\sqrt{(1+\eta_{-k_\parallel})(1+\eta_{k_\parallel})} - \sqrt{(1-\eta_{-k_\parallel})(1-\eta_{k_\parallel})}\right]^2}{4}$$

(3)

where $\eta_{k_\parallel}$ and $\eta_{-k_\parallel}$ are the radiation-directionality of the guided-resonance modes with opposite tangential wavevectors ($k_\parallel$ and $-k_\parallel$), which are the modes excited in the original scattering event and its time-reversal process, respectively. Two panels in Fig. 2b depict, respectively, how $R_{peak}$ and $R_{dip}$ vary against $\eta_{k_\parallel}$ and $\eta_{-k_\parallel}$. From the phase diagrams, we find that $R_{dip} = 0$ and $R_{peak} = 1$ can always be satisfied in the case of $\eta_{k_\parallel} = \eta_{-k_\parallel} = 0$, meaning that the metagrating has mirror symmetry (the red dot). Meanwhile, as the metagrating exhibits P-symmetry with $\eta_{k_\parallel} = -\eta_{-k_\parallel} = \eta$, we find that while $R_{dip} = 0$ is still guaranteed while $R_{peak}$ becomes a quadratic function of the directionality $R_{peak} = 1 - \eta^2$ (the blue dashed line). This reinforces our previous argument of realizing the desired spatio-spectral selectivity via designing a P-symmetric metagrating with an angle-dependent $\eta(\theta)$.

## 2.2 Design and fabrication of misaligned bilayer metagratings

We now consider a specific meta-structure that can exhibit a designable $\eta(\theta)$. As shown in Fig. 1c, the system consists of two identical silicon (n = 3.41) gratings separated by a spacer of thickness δ, immersed inside the background medium of $SiO_2$ (n = 1.444). Each grating consists of a periodic array (periodicity a = 613 nm) of silicon stripes with width $w_g$ = 425 nm and thickness $t_g$ = 210 nm, and two gratings exhibit a lateral misalignment Δg. We note that while a non-zero Δg breaks the mirror symmetry of the system, the P-symmetry is still maintained in such a configuration, making our system an ideal candidate to realize the spatio-spectral selectivity.

We now explicitly discuss how to design a meta-device to achieve spatio-spectral selectivity at normal incidence (see Supplementary Section 6 and Fig. S7 for discussions on oblique incidences). Since the P-symmetry ensures $R_{dip} = 0$ in our system, we only need to study the angle-dependent peak-reflection $R_{peak}(\theta) = 1 - (\eta(\theta))^2$. We find that two properties of the $\eta(\theta)$ function dictate the final performance of spatio-spectral selectivity. First, noting that $\eta(0°) \equiv 0$, we find that the edge angle $\theta_{edge}$ at which $|\eta(\theta_{edge})| = 0.894$ sets up a boundary of the high-$R_{peak}$ region (with $R_{peak} > 0.2$). Second, the angle bandwidth $\Delta\theta_{not}$ inside which $\eta$ maintains a high value ($|\eta(\theta)| \geq 0.894$) defines the region where the modes are *not* selected ($R_{peak} \leq 0.2$). Therefore, in order to achieve a better performance on angle selectivity, we need to make $\theta_{edge}$ smaller and $\Delta\theta_{not}$ larger.

Figures 2c and 2d depict, respectively, how $\theta_{edge}$ and $\Delta\theta_{not}$ vary against Δg and δ, calculated by finite-element-method (FEM) simulations on our metagratings with other structural parameters fixed. The overlapping between the blue region in Fig. 2c and the red region in Fig. 2d defines a parameter region to guide us design meta-gratings for generating high-performance spatio-spectral selectivity. Fixing δ = 35 nm by balancing device compactness and experimental feasibility, we choose two systems with Δg = 0.03a and Δg = 0.06a, denoted by the yellow triangle and the blue star in two phase diagrams, respectively, and numerically compare their performances. Upper panels in Figs. 2e and 2f depict the calculated wavelength dispersions of guided-resonance modes in these two systems, while lower panels show the calculated $\eta(\theta)$ functions of the $TE_1$ guided-resonance modes in different systems. Obviously, the device with Δg = 0.03a exhibits a poor spatio-spectral selectivity (see upper panel in Fig. 2e), caused by a large edge angle $\theta_{edge} = 21°$ and a small non-selective angular bandwidth $\Delta\theta_{not} = 4°$ (lower panel in Fig. 2e), in consistency with the phase diagrams shown in Figs. 2c and 2d. In sharp contrast, the device with $\Delta g = 0.06a$ exhibits a much better spatio-spectral selectivity, simply because the system exhibits a small edge angle of $\theta_{edge} = 10°$, together with a larger non-selective angular bandwidth of $\Delta\theta_{not} = 11°$.

We find that the underlying physics is governed by angle-dependent interferences between radiation fields from two single layers of the metagrating. Based on a Green's function method (Supplementary Section 4), we calculated the amplitude ratios and phase differences between fields radiated from two single grating layers to upward and downward directions in different systems. Figures. 2g and 2h compare, respectively, the angle-dependent amplitude ratios and phase differences of fields contributed by two single gratings radiating to two different directions. In case of $\theta = 0°$, we note that radiation fields from two single layers exhibit symmetry-protected properties: $A_1^{up}/A_2^{up} = A_2^{down}/A_1^{down}$, $\delta\varphi^{up} = -\delta\varphi^{down} + 2\pi$, resulting in symmetric radiation with $\eta \equiv 0$ independent of the misalignment $\Delta g$ (see Fig. S4). In the cases of $\theta \neq 0°$, however, such symmetry is broken and thus asymmetric radiation can appear. Interestingly, varying $\theta$ can dramatically modify the phase differences $\delta\varphi^{up}$ and $\delta\varphi^{down}$ in different ways. In particular, we get $\delta\varphi^{down} = \pi$ at $\theta = 18°$, leading to nearly perfect destructive interference for the downward radiation since the radiation amplitude ratio is close to 1. Noting that the same thing does not happen for the upward radiation, we thus get $\eta \approx 1$ at the vicinity of $\theta = 18°$. This well explains the key features of the $\eta \sim \theta$ curve shown in the lower panel of Fig. 2f for the system with $\Delta g = 0.06a$. Meanwhile, understanding the essential physics can guide us "design" the $\eta(\theta)$ function through tailoring the $\delta\varphi^{down} \sim \theta$ relation. Specifically, we need to design a system exhibiting such a $\delta\varphi^{down}(\theta)$ function: 1) $\delta\varphi^{down}$ should be close to $\pi$ at $\theta = 0°$ so that $\theta_{edge}$ can be small; 2) $\delta\varphi^{down}$ should pass through $\pi$ with a tiny slope so that $\Delta\theta_{not}$ can be large. Indeed, $\delta\varphi^{down}(\theta)$ of the optimized structure with $\Delta g = 0.06a$ satisfies the above two conditions (see Fig. 2h), while that of the un-optimized structure with $\Delta g = 0.03a$ does not (see Fig. S4), which explains why such a system does not exhibit the desired spatio-pectral selectivity (see Fig. S6).

To verify our theoretical predictions, we fabricate out a misaligned bilayer metagrating sample with $\Delta g = 0.06a$ and $\delta = 35$ nm using the electron-beam lithography technology. Among all fabrication processes, growing a flat and thin spacer with a thickness of 35 nm is crucial and the most challenging. Unfortunately, directly growing $SiO_2$ material onto the first-layer grating can inevitably lead to an uneven spacer, significantly degrading the device performance. To solve this issue, we first use hydrogen silsesquioxane polymers (HSQ) to fill in the gaps in the first-layer Si grating deposited on a 500 μm - thick fused silica substrate, and then etch away the HSQ above the grating to flatten the structure through modifying etching parameters and measuring the spacer thickness indirectly (Methods and Fig. S10). Following this flattening and thinning process, a 35 nm thick layer of $SiO_2$ is precisely deposited by inductively coupled plasma chemical vapor deposition (ICPCVD). The complete fabrication process can be found in Methods. Figure 3a shows a top-view optical image of the fabricated sample with a total footprint of 200 μm × 200 μm. Flat spacer with thickness of $\delta = 35$ nm and precise misalignment of $\Delta g = 37$ nm are unambiguously proven in the scanning electron microscope images depicted in Fig. 3b and 3c. Other structural parameters are $a = 613$ nm and $w_g = 425$ nm, in consistency with theoretical design. Additionally, we also put a 1 μm - thick HSQ layer onto the bilayer metagrating, to ensure that the whore structure is immersed inside the same background medium as required in our theory.

**2.3 Cross-polarization measurement of angle-contollable directionality**

With the fabricated sample at hand, we then utilize an angle-resolved micro-spectrometer system (ARMS, Angle-Resolved Micro-Spectrometer, ideaoptics, China) to characterize the radiation directionality of its guided-resonance modes excited at different angles. As we only care about the properties of the resonant modes, we employ a cross-polarization measurement method to separate the useful signals from the background signals including incident light and non-resonant scatterings[49]. In our experiment, we shine the sample by a s-polarized light beam but with **E** field rotated by 1° on the xy-plane, exhibiting the polarization component to excite the resonant modes (Methods and Fig. S9). Then, a p-polarized signal from the excited resonant modes can be extracted. By flipping the sample along the y-axis, we can measure upward and downward radiation signals correspondingly.

Left panels in Figs. 3d and 3e depict, respectively, the measured intensities of upward and downward signals radiated from our sample, as being excited by light beams with different wavelengths and incident angles. For TE$_1$ band that we consider, we find clearly that the upward radiation intensity $I_{up}$ is weaker than the downward counterpart $I_{down}$ in the angle range of $-25° < \theta < -5°$, consistent with simulation results (right panels in Figs. 3d and 3e). We next calculate the radiation directionality of the resonant mode via $\eta = \frac{I_{up} - I_{down}}{I_{up} + I_{down}}$ based on data obtained from measurements and simulations, and depict in Fig. 3f how the measured and simulated $\eta$ vary against $\theta$. We find that the measured $\eta$ (red dots) first decreases and then increases as $\theta$ varies from -25° to -5°, reaching its lowest value at -15°. The measured $\eta \sim \theta$ relation agrees well with the simulated curve (black line), both showing that the $\eta(\theta)$ function exhibiting the desired angle-controlled property for achieving spatio-spectral selectivity. We note that the $\eta \sim \theta$ relation retrieved from simulations (solid line in Fig. 3f) deviate slightly from that calculated by quasi-normal-mode method (lower panel in Fig. 2f), since the former approach could involve coupling between TE$_1$ mode and TM$_2$ mode especially in the small-$\theta$ region.

### 2.4 Demonstration of spatio-spectral selectivity

Now that our sample exhibits the desired $\eta(\theta)$ function, we continue to experimentally demonstrate that it possesses spatio-spectral selectivity at $\theta = 0°$, as predicted in Fig. 2. We first measure the reflection spectra of our system under different incident angles, and then depict in Fig. 4a the obtained reflectance versus wavelength and incident angle. Figure 4a shows clearly that our device exhibits high reflectivity at the vicinities of $\theta = 0°$ and $\lambda = 1349$ nm with an angular bandwidth of 10° and a spectral bandwidth of 14 nm, well illustrating the spatio-spectral selectivity as desired. These results agree well with the simulation results shown in Fig. 4b, with the latter exhibiting a better performance in term of both angular bandwidth (8°) and wavelength bandwidth (5 nm).

Figure 4c compares the measured and simulated reflection spectra of our sample at six different incident angles. Both simulations and experiments reveal that the peak reflectance $R_{peak}$ of our sample first decreases and then increases as $\theta$ increases, creating a (nearly)-zero-reflection region in $10° < \theta < 20°$ induced by high radiation-directionality of the resonant modes. Noting that $R_{peak}$ under normal incidence is guaranteed to be 100% in theory protected by the P-symmetry of our system. Such peculiar behaviors help to pick out a single mode in the angle-wavelength map, as shown in Figs. 4a and 4b. However, compared to simulations, experimental value of $R_{peak}$ at normal incidence (~0.47) is significantly smaller than 100% and all reflection peaks in experimental Fano spectra exhibit relatively lower Q factors, which can be attributed to sample imperfections inevitably existing in fabrications.

Finally, we experimentally demonstrate a practical application of our device. We first fabricate a bilayer metagrating with the same geometrical parameters as in Fig. 3-4 but with a larger size of 1 mm × 1 mm. We next put our metagrating in front of a mask containing a cartoon architecture pattern, and then illuminate the metagrating by a wavelength-tunable laser beam. Via tuning the orientation angle of our metagrating with respect to the light incidence direction, we can effectively filter out light components with desired propagating directions and wavelengths, which in turn, serve as the light source to illuminate our mask. We first perform a benchmark test with the metagrating taken away, and find that the target pattern can be captured on the detecting plane at all wavelengths of interests (Supplementary Section 9 and Fig. S11). When the metagrating is inserted and the incident angle is fixed at 0° (see Fig. 5a), we find that the recorded pattern exhibits a pronounced darkening only at the wavelength of 1342 nm, obviously caused by the spatio-spectral selectivity of our metagrating. In contrast, as we change the incident angle to 10° to repeat our experiments, we find that the pattern can be correctly captured at all wavelengths of interests (Fig. 5b), since all wavelength components are allowed to pass through the metagrating without significant losses under oblique incidence. We further examine the performance of such selective imaging within a narrower wavelength bandwidth (Fig. 5c) or a narrower angular bandwidth (Fig. 5d). Figure 5c shows that the brightness of the imaging pattern is obviously diminished as $\lambda$ lies inside $1342 \pm 5$ nm, as the incident angle is fixed at 0°. Meanwhile, fixing $\lambda = 1342$ nm and varying the incident angle $\theta$, we find from Fig. 5d that the imaging pattern turns bright as $|\theta| = 4°$. All these results confirm that our misaligned bilayer metagrating can achieve narrow-angle and narrow-wavelength selective imaging at 1342 nm and 0°.

## 3 Discussion and Conclusion

In short, we propose a generic method to design photonic structures exhibiting spatio-spectral selectivity, and experimentally verify the concept in the near-infrared. We show that the radiation asymmetry of a resonant mode in a photonic structure plays a vital role to manipulate the Fano-resonant intensity of the system, and establish a complete phase diagram to guide designing optimized structures achieving high-performance spatio-spectral selectivity. We find that misaligned bilayer metagratings support resonant modes with angle-dependent radiation directionality, and design an optimized metagrating to achieve spatio-spectra selectivity guided by the phase diagram. We fabricate out the sample according to the design, and experimentally characterize its optical properties. Our experiments reveal that the meta-device exhibits high reflectivity only for light with 1349 nm wavelength and at normal incidence, exhibiting the desired spatio-spectral selecticity with a narrow angular width (±5°) and wavelength width (14 nm). Finally, we experimentally demonstrate spatio-spectral selective imaging with our fabricated bilayer meta-grating, showing narrow-band and high-contrast performances. Apart from the proposed mechanism that is different from all previous ones, our realizations of spatio-spectral selectivity can find many applications in practice, such as coherent thermal emission[50], high-beam-quality lasing[5] and multifunctional glasses in augmented reality[33,40]. Moreover, fabricating bilayer systems with precision control can also help studying other interesting phases such as moiré photonics[49,51] and topological photonic crystals[52,53].

## 4 Methods

### 4.1 Numerical simulations

Eigen modes of bilayer metagratings are solved by COMSOL Multiphysics software. Two-dimensional model with axis of x and z is created for simulation. Bloch boundaries are used in the x direction and perfect matched layers are applied in the z direction. The radiation intensities of eigen modes in upward/downward direction are calculated by integrating the electric field in the upper/lower plane out of the bilayer metagrating. And the directionality of eigen modes are evaluated by the radiation intensity on both sides. As for the angle-resolved reflection spectrum, we adopt the rigorous coupled-mode theory for simulating the scattering of periodic structures.

### 4.2 Sample fabrication

The proposed bilayer metagratings are fabricated through a series of processes, including inductively coupled plasma chemical vapor deposition (ICPCVD, Oxford PlasmaPro System 100ICP180-CVD), electron-beam lithography (EBL, Raith Vistec EBPG-5000plusES), inductively coupled plasma etching (ICP, Oxford PlasmaPro System 100ICP180) and reactive ion etching (RIE, Oxford PlasmaPro System 100RIE180). Fig. S10 illustrates the step-by-step fabrication procedure.

To prepare for the well-aligned bilayer metagrating, a pair of gold (Au) alignment markers are patterned by EBL, followed by Au evaporation and lift-off process. A 210 nm thick layer of amorphous silicon (α-Si) is deposited on a 500 μm thick fused silica substrate by ICPCVD (Fig. S10a). Subsequently, a 200 nm thick positive electron beam resist (ARP6200) is spin-coated on the α-Si film at 4000 rpm, and the sample is baked on a hot plate at 183°C for 10 minutes (Fig. S10b). To mitigate charging effects during the EBL process, a 30 nm thick aluminum layer is thermally evaporated onto the resist. Next, a grating pattern is defined in the resist by EBL system at an acceleration voltage of 100 keV (Fig. S10c). The patterns are aligned by recognizing the Au markers under the EBL. The aluminum layer is then removed using phosphoric acid, followed by the development of ARP6200 with dimethylbenzene. Next, the 210 nm thick α-Si layer is etched by ICP, and any residual resist is removed through $O_2$ plasma descum to form the first-layer metagrating structure (Fig. S10d).

Before depositing a 35 nm thick silicon dioxide layer, we first use the hydrogen silsesquioxane polymers (HSQ) to fabricate the flat filled layer. A 1000 nm thick HSQ is spin-coated onto the α-Si layer, covering both the patterned metagrating area and the unpatterned film area, followed by baking at 300°C for 6 hours (Fig. S810e). Then multiple etching steps were followed. In this process, the primary challenge is determining when the HSQ above the metagrating area has been exactly etched away. Note that we cannot directly measure the HSQ thickness above the metagrating area via conventional methods such as a film thickness tester. Instead, we measured the HSQ thickness above the unpatterned film surrounding the metagratings according to the reflection spectrum and measure the thickness difference by the surface profiler. These two values then reveal the actual thickness of HSQ on the first-layer metagrating. Since the thickness difference changes with etching time, multiple etching steps were needed, in which RIE etching and thickness measurement were alternately performed. After this complex process, we finally etched the HSQ above the metagrating as close to 0 as possible (Fig. S10f). Thanks to the flattening process above, a 35 nm even silicon dioxide (SiO2) layer was then precisely deposited by ICPCVD as a spacer layer (Fig. S10g).

Furthermore, a 210 nm layer of α-Si was deposited by ICPCVD for etching the second layer metagrating (Fig. S10h). The second-layer metagrating was fabricated using the same method as the first layer (Figs. S10i-k). During EBL exposure, the Au markers aided in the alignment of the nanostructures on the second layer. Finally, another round of HSQ spin-coating and baking was performed to create the top cladding of the bilayer metagrating structure (Fig. S10l).

### 4.3 Cross-polarization measurement for directionality

The measurement of upward and downward radiation intensities is based on an angle-resolved micro-spectrometer system (ARMS, Angle-Resolved Micro-Spectrometer, ideaoptics, China). The polarization of the incident light is set along the y-direction (s-polarization), and the reflected (transmitted) light is collected by a 100X near-infrared microscope objective with a numerical aperture of 0.85, with the detection area limited to 60 μm × 60 μm. The polarization at the detection end is set along the x-direction (p-polarization) (Fig. S9a).

The measurement expressions for upward and downward radiation intensities and directionality are as follows:

$$I_{up}(\theta,\lambda) = \frac{\text{SAR}(\theta,\lambda)}{\chi(\theta,\lambda) \cdot \text{LSR}(\theta,\lambda)}$$

$$I_{down}(\theta,\lambda) = \frac{\text{SAT}(\theta,\lambda)}{\chi(\theta,\lambda) \cdot \text{LST}(\theta,\lambda)}$$

$$\eta(\theta,\lambda) = \frac{I_{up}(\theta,\lambda) - I_{down}(\theta,\lambda)}{I_{up}(\theta,\lambda) + I_{down}(\theta,\lambda)}$$

where SAR (SAT) represents the cross-polarized angular reflection (transmission) spectrum obtained from the sample. During the measurement of SAR, the bilayer metagratings sample is rotated by 1° in the xy-plane. For the measurement of SAT, the bilayer metagratings sample is flipped along the y-axis and then rotated by -1° in the xy-plane. LSR (LST) represents the reference wavelength-angle reflection (transmission) spectrum measured. During the measurement of LSR (LST), the polarization of the incident light is set to s-polarization, and no polarization is set at the detection end. To enhance the signal intensity of upward (downward) measurements and prevent CCD overexposure during LSR (LST) measurements, a neutral density filter with an attenuation rate of $\chi$ is placed at the detection end. It is noteworthy that the same neutral density filter is used, ensuring the comparability of upward and downward radiation intensities, thereby eliminating the influence of the filter when calculating directionality.

### 4.4 Spatio-spectral selectivity spectral measurement

The measurement of spatio-spectral selectivity spectral is based on ARMS. The polarization of the incident light is set along the y-direction (s-polarization), and the reflected light is collected by a 100X near-infrared microscope objective with a numerical aperture of 0.85. The detection area is limited to 60 μm × 60 μm.

### 4.5 Spatio-spectral selective imaging characterization

In the experiment (Fig. S11b), the sample is positioned at the back focal plane of the first lens (Lens 1). Different from spectral measurement, here we fabricate a larger sample (1 mm × 1 mm) for imaging characterization, which is more compatible to practical application. The focused spot is converted into a collimated beam by an objective lens (Lens 2) and then illuminated onto a mask (there is a cartoon architecture pattern on the mask). The transmission pattern is captured by an imaging system behind the mask, which consists of an objective lens (Lens 3), a beam splitter and a CCD camera. A halogen lamp is employed to assist in confirming the position of the pattern.

**Acknowledgments:** We thank Lei Shi and Jiajun Wang for discussions and experimental assistance.

**Funding:**

    National Natural Science Foundation of China grant 62035016

    National Natural Science Foundation of China grant 12221004

    National Natural Science Foundation of China grant 62192771

    National Key Research Development Program of China grant 2021YFB2802300

    National Key Research Development Program of China grant 2022YFA1404304

    National Key Research Development Program of China grant 2022YFA1404700

    National Key Research Development Program of China grant 2023YFB2806800

    Guangdong Basic and Applied Basic Research Foundation grant 2023B1515040023

    Natural Science Foundation of Shanghai grant 23dz2260100

**Author contributions:**

    Conceptualization: Z.P.Z., J.W.D., L.Z.

    Methodology: Z.P.Z, X. Z., H.L.Z., M.Y.L., Z.M.C., J.W.D.

    Investigation: Z.P.Z, X. Z., H.L.Z., M.Y.L., Z.M.C., X.T.H., X.D.C.

    Visualization: Z.P.Z, X. Z., H.L.Z., J.W. D.

    Funding acquisition: X.T.H., X.D.C., J.W. D., L.Z.

    Project administration: J.W.D.

    Supervision: J.W.D., L.Z.

    Writing – original draft: Z.P.Z, X. Z., H.L.Z., X.D.C., J.W.D.

    Writing – review & editing: Z.P.Z, X. Z., H.L.Z., X.D.C., J.W.D., L.Z.

**Competing interests:** Authors declare that they have no competing interests.

**Availability of data and materials:** All data are available in the main text or the supplementary materials. Additional data are available from the corresponding authors upon reasonable request.

**Ethics approval and consent to participate**

    Not applicable.

**Consent for publication**

    Not applicable.

**Additional information**

    Supplementary information

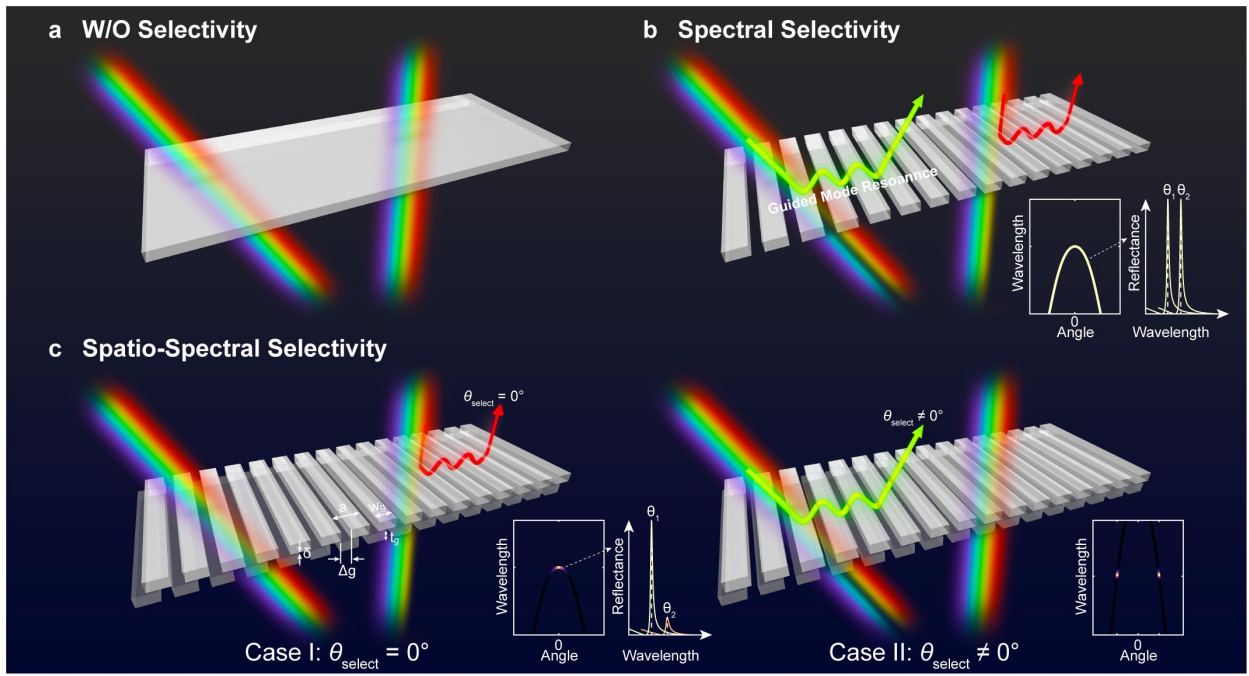

**Fig. 1. Concept of spatio-spectral selectivity enabled by symmetry and directionality design.
a,** A slab shows non-selectivity for wavelength and angle, leading to transmission of image without filtering any spatio-spectral information. **b,** A single layer metagrating shows an angle-dependent spectral selectivity, resulting in only filtering spectral information of the image. Inset exhibits the dispersion of the guided mode resonance. Totally reflection happens along the dispersion line. **c,** A misaligned bilayer metagrating shows spatio-spectral selectivity at normal incidence (left, Case I: $\theta_{select} = 0$) and oblique incidence (right, Case II: $\theta_{select} \neq 0$), in which only specific angle and wavelength is filtered. Inset shows the angle-dependent reflection intensity of the guided mode resonance.

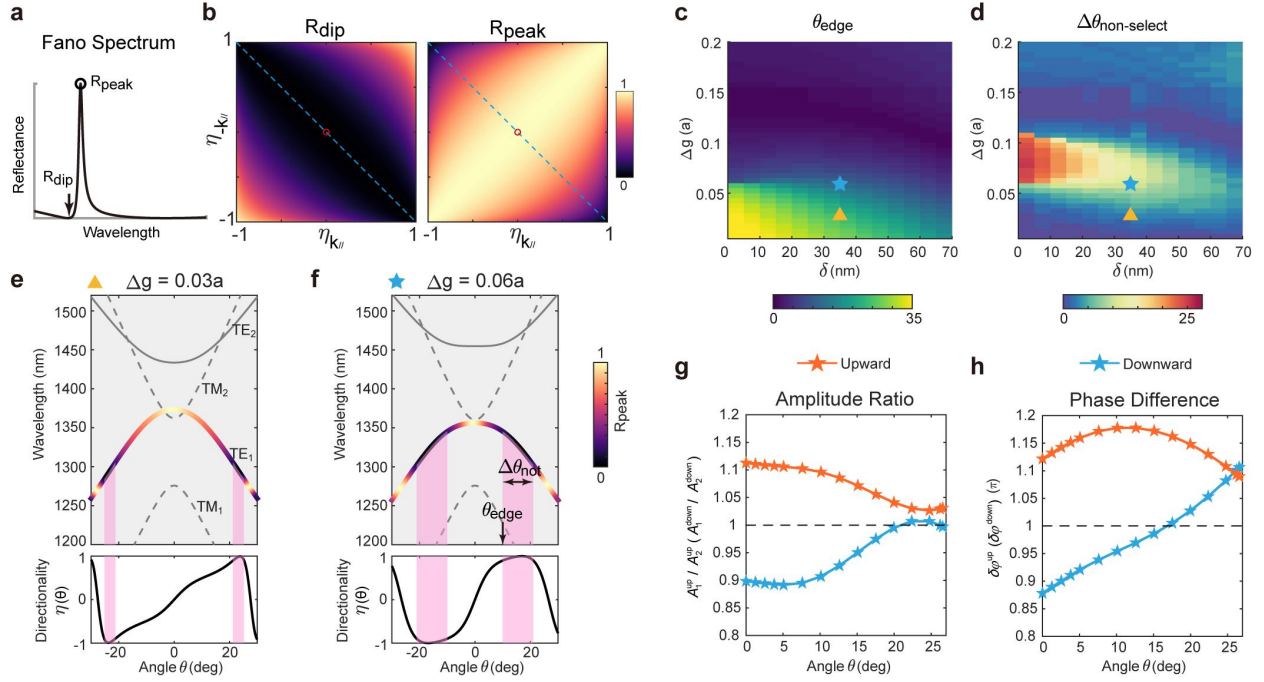

**Fig. 2. Design of misaligned bilayer metagratings with high directionality in wide angle. a,** Fano spectrum in reflection induced by the guided mode resonance, which has a minimum value ($R_{dip}$) and maximum value ($R_{peak}$) of reflectance. **b,** The reflection dip and peak of Fano spectrum as functions of directionality of guided-resonance modes with matched in-plane momentum $k_{//}$ and $-k_{//}$. The red dots correspond to the case of mirror symmetry and dashed lines represent the case of P symmetry. **c,d,** The edge angle of high-directionality region $\theta_{edge}$ (c) and angular bandwidth of low-directionality region $\Delta\theta_{non\text{-}select}$ (d) for $TE_1$ band under different spacer thickness $\delta$ and misalignment $\Delta g$. **e,f,** (Top) Band structure of bilayer metagratings with $\Delta g = 0.03a$ and $0.06a$, corresponding to blue star and yellow triangle marker in (c) and (d), respectively. The color of $TE_1$ band indicates the reflection peak value. (Bottom) Corresponding directionality as a function of angles for $TE_1$ band. The pink areas indicate the reflection peak is lower than 0.2. **g,h,** Calculated radiation amplitude ratio (g) and phase difference (h) between upper and lower metagratings in upward (orange) and downward (blue) directions.

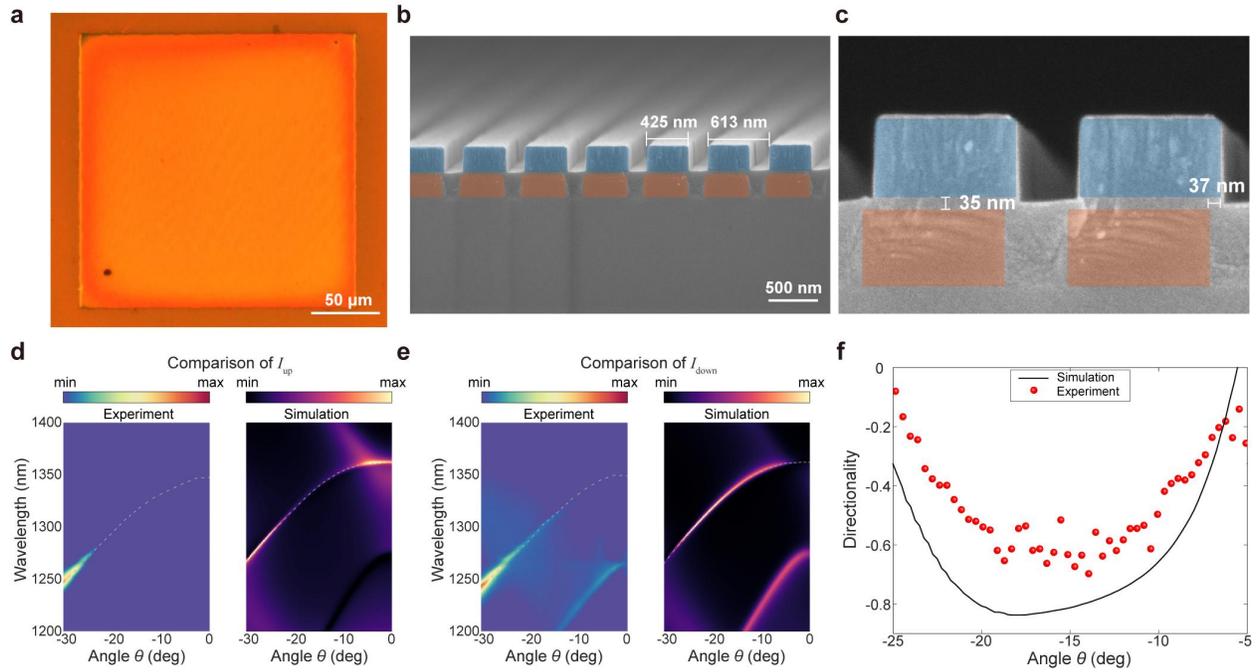

**Fig. 3. Cross-polarization measurement for directionality along with incident angles. a,** Optical microscope image of the misaligned bilayer metagratings with a footprint of 200 μm by 200 μm. **b,c,** Scanning electron microscope images of the fabricated bilayer metagratings sample (without cladding) from tilted (**b**) and side (**c**) views. Pseudo-coloring is used to mark the position of the gratings. **d,e,** Experimental and simulated results of the radiation intensity ($I_{up}$ and $I_{down}$) of the misaligned bilayer metagratings with varying incident light wavelengths and angles. The incident light is s-polarized, and the receiving end is p-polarized. **f,** Results of the variation of directionality of the misaligned bilayer metagratings with incident angles. The red colored dots represent experimental results, and the black curve represents simulation results. The measurement data correspond to the incident angle variation along the x-z plane of the sample.

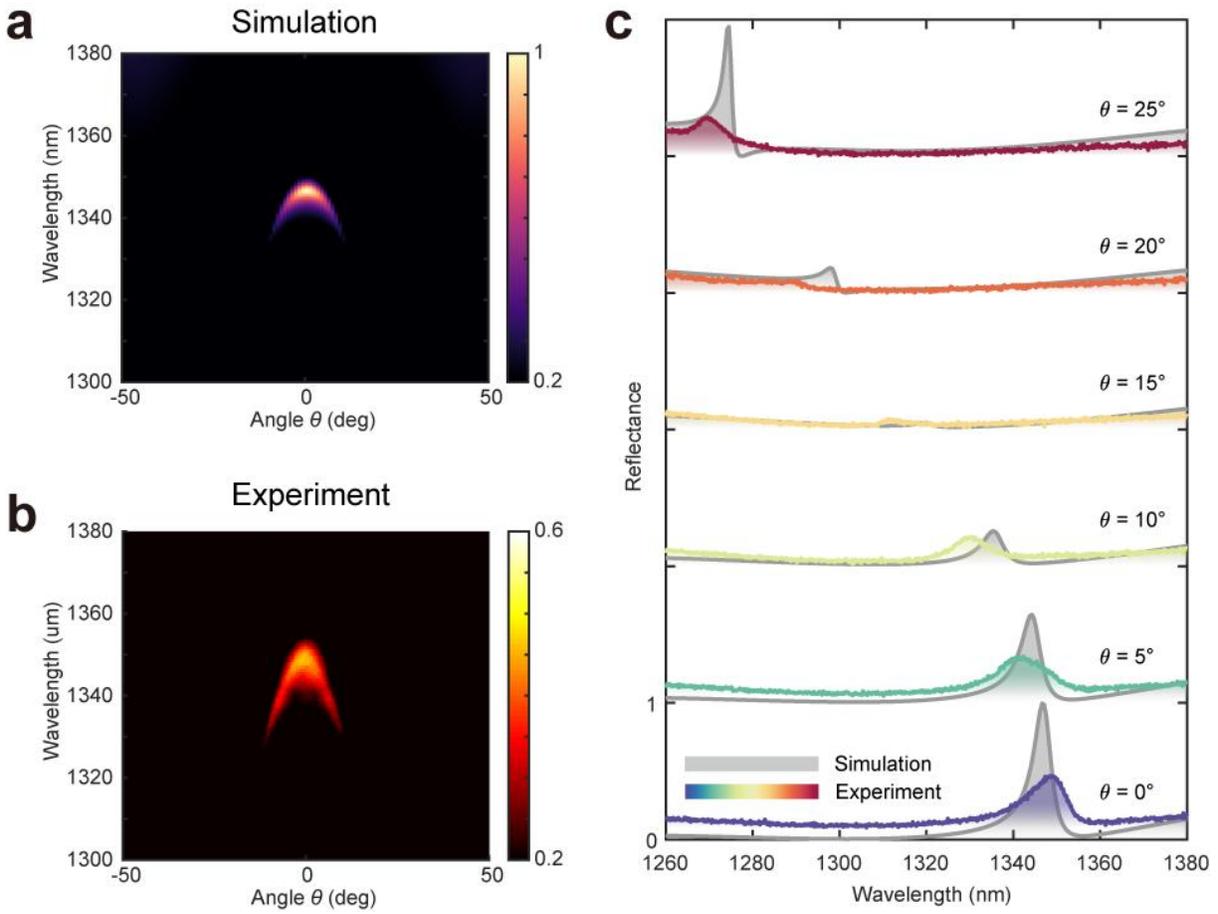

**Fig. 4. Observation of spatio-spectral selectivity. a,b,** Angle-resolved reflection spectrum in simulation (**a**) and experiment (**b**). **c,** Comparison of reflection spectrum between simulation and experiment at incident angles of 0°, 5°, 10°, 15°, 20° and 25°. Low reflection can be observed in angles ranging from 10° to 20°, which is related to high directionality of the resonant modes in these angles, demonstrated in Fig. 3.

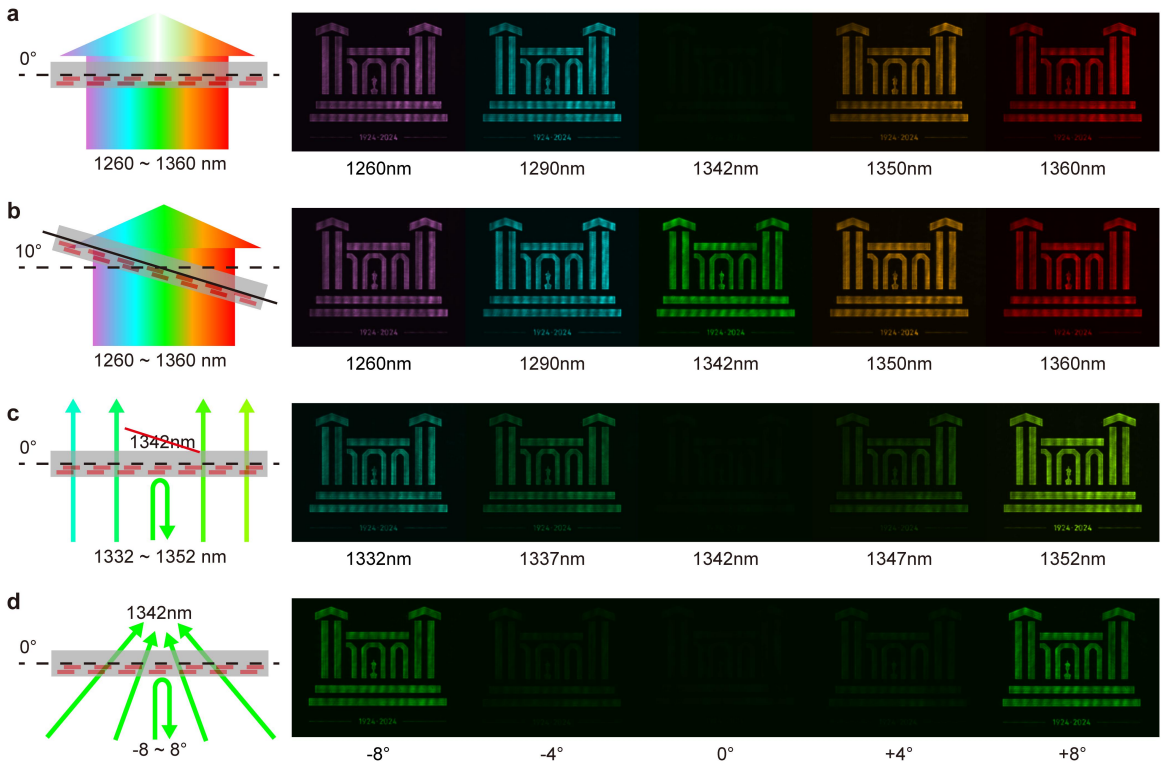

**Fig. 5. Spatio-spectral selective imaging performance of misaligned bilayer metagratings. a,** Imaging performance under normal incidence within a broadband range from 1260 nm to 1360 nm. Light only at 1342 nm is reflected, while other wavelengths go through the sample without significant loss. **b,** Imaging performance under oblique incidence (10°) within a broadband range from 1260 nm to 1360 nm. All wavelengths are allowed to pass through. **c,** Imaging performance under normal incidence within a narrow band range from 1332 nm to 1352 nm. Transmission at 1342 nm is suppressed, while other wavelengths transmit normally. **d,** Transmission imaging performance at different angles (-8° to +8°). Light is blocked from passing only at 0°, while it is normal at other angles.

23